\documentstyle{article}
\newcommand{\newsection}{    
\setcounter{equation}{0}
\section}
\def\appendix#1{
\addtocounter{section}{1}
\setcounter{equation}{0}
\renewcommand{\thesection}{\Alph{section}}
\section*{Appendix \thesection\protect\indent #1}
\addcontentsline{toc}{section}{Appendix \thesection\ \ \ #1}
}

\newcommand{\ov}[1]{\overline{#1}}

\def\be{\begin{equation}}
\def\ee{\end{equation}}
\newcommand{\beq}{\begin{equation}}
\newcommand{\eeq}{\end{equation}}
\newcommand{\bea}{\begin{eqnarray}}
\newcommand{\eea}{\end{eqnarray}}

\newcommand{\th}{\theta}

\newcommand{\vph}{\varphi}

\newcommand{\eps}{\varepsilon}

\def\e{{\,\rm e}\,}

\renewcommand{\d}{{{\partial}}}

\begin{document}
\topmargin 0pt
\oddsidemargin 5mm
\headheight 0pt
\headsep 0pt
\topskip 9mm

{{}\hfill SMI-Th--99/04}

\begin{center}
\vspace{26pt}
{\large \bf {AdS$_3$/CFT$_2$ correspondence at finite temperature}}
\vspace{26pt}

{\sl L.\ Chekhov}
\footnote{email: chekhov@mi.ras.ru}
\\
\vspace{6pt}
Steklov Mathematical Institute\\ Gubkin
st.\ 8, 117966, GSP-1, Moscow, Russia\\

\end{center}
\vspace{20pt}
\begin{center}
{\bf Abstract}
\end{center}

\noindent
The AdS/CFT correspondence is established for
the AdS$_3$ space
compactified on a solid torus with the CFT field on
the boundary. Correlation functions that correspond to the bulk
theory at finite temperature are obtained in the regularization
a'la Gubser, Klebanov, and Polyakov. 
The BTZ black hole solutions
in AdS$_3$ are T-dual to the 
solution in the AdS$_3$ space without singularity.


\newsection{Introduction}
The AdS/CFT correspondence~\cite{Maldacena,Polyakov,Witten}
has been verified 
for interacting field cases~\cite{inter,HF}
(three- and four-graviton scattering, etc.)
and it is interesting to check it also in
cases where the space--time geometry is more involved than
the spherical one. Various approaches to this problem were proposed 
in~\cite{Bonelli} and~\cite{Ch1}.

In~\cite{Bonelli}, it was
proposed, starting from a two-dimensional (compact) manifold $M$,
to consider a theory on the space $M\times {\bf R}_+$ endowed with the
AdS metric. From the topological standpoint, this, however, results
in a singularity as $s\in{\bf R}\to+\infty$ (because $M$ is not
necessarily simply connected), and one must impose an additional
condition on the fields of the theory (the fast decrasing at infinity)
in order to make the field configuration smooth.

In~\cite{Ch1}, we considered the massless scalar
field theory on AdS$_3$ space.
(The method can be generalized to the massive modes of
AdS$_n\times S^{d+1}$ space.) In this paper, we also consider the case
of a general (not necessarily rectangular)
torus and discuss the physical content of the results obttained.
We also consider the case of a homogeneously
compactified AdS$_3$ manifold without (topological) singularity in the
interior (as there is no boundary as $s\to\infty$; 
our calculations technically resemble the bulk calculations of~\cite{HF}).
We show that the classical scalar field theory on the AdS$_3$ manifold
in the bulk gives us the appropriate quantum correlators on the boundary.

The compactification corresponds to considering
the finite temperature case, so we calculate
correlators of boundary fields
for the AdS$_3$ space at finite temperature~\cite{Wit2}.
Following~\cite{Wit2}, one must take into account all possible solutions of
the Einstein gravity that have the anti-de~Sitter metric at infinity.
In the AdS$_3$ case, however, the black-hole solutions of Hawking and
Page~\cite{HP83,Wit2} turn out~\cite{CT,MS} to be T-dual to the
case of a pure AdS$_3$ space without internal singularities;
this holds for the corresponding correlation functions as well.

In Sec.~2, we recall the general structure of AdS$_3$ manifolds and introduce
the $\eps$-cone regularization in order to make the volume and the 
boundary area finite. In Sec.~3, we review the relationship between
the black hole and empty space solutions of the AdS$_3$ gravity.
In Secs.~4 and~5, we formulate the scalar field theory in AdS$_3$, while the
corresponding correlation functions are calculated by using an improved
technique in Sec.~6. Separate investigation is performed for the case
where a conical singularity (the angle deficit) is developed in the
vicinity of a (unique) closed geodesic inside the solid torus; 
the correlation functions are found for the case where the angle
deficit is a rational number (Sec.~7). A brief discussion is in Sec.~8.

\newsection{Geometry of AdS$_3$ manifolds}

The group $SL(2,{\bf C})$ of conformal transformations of the complex
plane admits the continuation to the upper half-space ${\bf H}_3^+$
endowed with the constant negative curvature (AdS$_3$ space).
In the Schottky uniformization picture, Riemann surfaces of higher genera
can be obtained from ${\bf C}$ by factorizing it over a finitely
generated free-acting discrete subgroup $\Gamma\subset SL(2,{\bf C})$.
Therefore, we can continue the action of this subgroup to the
whole AdS$_3$ and, after factorization, obtain a three-dimensional manifold
of constant negative curvature (an AdS$_3$ manifold)
whose boundary is (topologically)
a two-dimensional Riemann surface~\cite{Manin,Sel}.

We consider the simplest case of a genus one AdS$_3$ manifold
obtained after the identification
\be
(w,{\ov w},s)\sim (qw,{\ov q}{\ov w},s|q|),
\label{a1}
\ee
where $q=\e^{a+ib}$ is the modular parameter, $a,b\in {\bf R},\ a>0$, \ 
$w,\ov w=x+iy,\,x-iy$ are
the coordinates on $\bf C$, and $s$ is the third coordinate on AdS$_3$.

Adopting the AdS/CFT correspondence approach, we should first regularize
expressions in order to make them finite. In the AdS$_n$ case with the
brane singularity
at infinity, this was done by setting the
boundary data on the $\eps$-plane~\cite{Polyakov}
rather than at infinity (zero plane).
However, in our case, we cannot take an $\eps$-plane because it is not
invariant w.r.t.\ transformations (\ref{a1}). Instead, we can set
the boundary data on the $\eps$-{\it cone}---the set of points
$$
z/r=\eps, \quad r^2=w\ov w+s^2.
$$
Given the boundary data on this cone, we
fix the problem setting---the Laplace equation then has a unique solution
(the Dirichlet problem on a compact manifold).

Geometrically, performing the $\eps$-cone
regularization and factorizing over
the group $\Gamma$ of transformations (\ref{a1}), we obtain the solid
torus on whose boundary (the two-dimensional 
torus) the CFT fields dwell. The ``center'' of the torus is 
a unique closed geodesic, which has the length $\log|q|$
(the image of the vertical half-line $z=\ov z=0$),
while the AdS-invariant
distance $r$ from this geodesic to the image of the $\eps$-cone is constant,
$\cosh r=1/\eps$.

\newsection{Black hole solution in AdS$_3$ in spherical coordinates}

In order to operate with the cone geometry, it is convenient to
reformulate the free-field problem on
AdS$_3$ in the spherical coordinates $r\equiv\e^\tau,\th,\vph$~\cite{CT}
in which relations (\ref{a1}) read
\bea
&{}&(\tau,\th,\vph)\sim (\tau+ma, \th, \vph+mb+2\pi n),\qquad m,n\in \bf Z,
\label{a2}
\\
&{}&dl^2=\bigl(d\tau^2+d\th^2+\sin^2\th\,d\vph^2\bigr)/\cos^2\th,
\nonumber
\eea
where $dl^2$ is the invariant interval squared.
On the cone $\sin\th=\eps$, the usual toric periodic conditions
are imposed on fields depending on the two-dimensional variables~$\tau$ 
and~$\vph$.

We prefer to work in the standard spherical coordinates 
$s=r\cos\th$, \ $y=r\sin\th\sin\vph$, and $x=r\sin\th\cos\vph$.
Upon the identification
\be
\rho=b\tan\theta,\quad t=b\log r,\quad \vph=\vph
\ee
we obtain the AdS$_3$ metric in the standard form~\cite{HP83},
\be
ds^2=\left(1+\frac{\rho^2}{b^2}\right)dt^2+\frac{1}{1+\rho^2/b^2}d\rho^2
+\rho^2d\vph^2,
\ee
with fields to be regular at $\rho=0$.

The remarkable fact is that the {\it spinning\/} black hole solution
metric (with the nonzero angular momentum~$J$ and mass~$M$) 
given by the interval
\bea
&{}&ds^2=(N^\perp)^2d\rho^2+(N^\perp)^{-2}dr^2+r^2(d\phi+N^\phi d\rho)^2,
\nonumber\\
&{}&N^\perp=\left(-M+\frac{r^2}{l^2}-\frac{J^2}{4r^2}\right)^{1/2},\qquad
N^\phi=-\frac{J}{2r^2},
\label{BH1}
\eea
with the following indentifications imposed:
\bea
&{}&\phi\sim\phi+2\pi,\qquad \left({\rho\atop\phi}\right)\sim
\left({\rho\atop\phi}\right)+\frac{2\pi}M\left({r_+\atop |r_-|/l}\right),
\nonumber\\&{}&
r_\pm^2=\frac{Ml^2}2\left[1\pm\left(1+\frac{J^2}{M^2l^2}\right)^{1/2}\right],
\eea
can be reduced by the coordinate transformations
\be
\vph=\frac{r_+}{l^2}\rho+\frac{|r_-|}{l}\phi,\quad
\tau=\frac{r_+}{l}\phi-\frac{|r_-|}{l^2}\rho,\quad
\th=\arcsin\sqrt{\frac{r^2-r_+^2}{r^2-r_-^2}}
\label{BH2}
\ee
to coordinates (\ref{a2}) with $a=2\pi r_+/l$ and $b=2\pi|r_-|/l$~\cite{CT}.

Therefore, metric (\ref{BH1}) with the respective (two-dimensional)
radial and angular coordinates~$\rho$ and~$\phi$ becomes metric (\ref{a2})
with the respective coordinates~$\tau$ and~$\vph$. Note that (for $J=0$)
radial and angular coordinates readily interchange and we obtain the torus
that is $T$-dual to the initial one.

\newsection{Scalar field on AdS$_3$ in spherical coordinates}

The action of the massless scalar field~$\Phi$ on AdS$_3$
in coordinates (\ref{a2}) is
\bea
&{}&\int\frac{dx\,dy\,ds}{s^3}\bigl\{s^2\d_s\Phi\d_s\Phi+s^2(\d_x\Phi\d_x\Phi
+\d_y\Phi\d_y\Phi)\bigr\}=\nonumber\\
&{}&\qquad=
\int\,\tan\th d\tau\,d\th\,d\vph\left\{\d_\tau\Phi\d_\tau\Phi
+\d_\th\Phi\d_\th\Phi
+\frac{1}{\sin^2\th}\d_\vph\Phi\d_\vph\Phi\right\}
\label{action}
\eea
It admits the variable separation,
\be
\Phi(\tau,\th,\vph)=
\sum_{k,m\in {\bf Z}}^{}\Phi_{k,m}(\tau)Y_{k,m}(\sin\th)X_m(\vph).
\ee
Taking into account 
the periodicity conditions
\be
\Phi(\tau+\log|q|,\th,\vph+\arg q)=\Phi(\tau,\th,\vph),
\qquad
\Phi(\tau,\th,\vph+2\pi)=\Phi(\tau,\th,\vph),
\label{e3}
\ee
where $q$ is the modular parameter of the torus,
\be
q=\e^{a+ib},\quad a,b\in{\bf R},
\label{e4}
\ee
we obtain
\bea
&{}&X_m(\vph)=\e^{im\vph}, \qquad \d_\vph^2X_m=-m^2X_m;
\label{vph}\\
&{}&\Phi_{k,m}(\tau)=
\e^{i\bigl[-\frac{mb}{a}+\frac{2\pi k}{a}\bigr]\tau},
\qquad \d_\tau^2 \Phi_{k,m}(\tau)=-\frac{w^2}{4}\Phi_{k,m}(\tau),\\
&{}&w\equiv -\frac{mb}{2a}+\frac{\pi k}{a}.
\label{ww}
\eea
For the function $Y_{k,m}(\sin\th)$, the eigenvalue problem is
\be
\frac{\cos\th}{\sin\th}\d_\th\left(\frac{\sin\th}{\cos\th}
\d_\th Y_{k,m}(\sin\th)\right) -\frac{m^2}{\sin^2\th}Y_{k,m}(\sin\th)
=\frac{w^2}{4} Y_{k,m}(\sin\th).
\ee
We substitute $\rho$ for $\sin\th$ and consider the problem on the
interval $1-\eps\ge\rho\ge0$ with the regularity condition at $\rho=0$,
\be
Y''_{k,m}(\rho)+\frac{1}{\rho}Y'_{k,m}(\rho)
-\frac{m^2}{\rho^2(1-\rho^2)}Y_{k,m}(\rho)=\frac{w^2/4}{1-\rho^2}
Y_{k,m}(\rho).
\label{e1}
\ee

Equation (\ref{e1}) can
be reduced to the standard hypergeometric equation whose
general solution that is regular at $\th=0$ yields
\bea
\Phi(\tau,\th,\vph)&=&\sum_{m,k\in{\bf Z}}^{}\e^{im\vph}\,
\e^{i\bigl[-\frac{mb}{a}+\frac{2\pi k}{a}\bigr]\tau}
\times C_{k,m}\times Y_{k,m}(\sin\th),
\nonumber\\
Y_{k,m}(\sin\th)&=&[\sin\th]^{|m|}\,
{}_2F_1\left(\frac{|m|}{2}+i\frac{-mb+2\pi k}{2a},
\frac{|m|}{2}-i\frac{-mb+2\pi k}{2a}; |m|+1; \sin^2\th\right),
\label{e5}
\eea
where ${}_2F_1(a,b;c;z)$ is the hypergeometric series,
$$
{}_2F_1(a,b;c;z)\equiv\sum_{k=0}^{\infty}\frac{(a)_k(b)_k}{(c)_k\cdot k!}z^k,
\qquad (a)_k\equiv\prod_{i=0}^{k-1}(a+i),
$$
and $C_{k,m}$ are the mode amplitudes.

Expression (\ref{e5}) is singular at $z=1$ and we must find its asymptotic
behavior for $z=1-\eps$. 
For $c-a-b\in{\bf Z}$, the
exact relation is (see, e.g., formula 7.3.1.31 from~\cite{BPM})
\bea
\Bigl.{}_2F_1(a,b;c;z)\Bigr|_{c=a+b+m}&=&
\Gamma\left[\begin{array}{ll} m,& a+b+m\\
                              a+m,& b+m \end{array}\right]
\sum_{k=0}^{m-1}\frac{(a)_k(b)_k}{k!(1-m)_k}(1-z)^k-
\nonumber\\
&{}&-\Gamma\left[\begin{array}{c} a+b+m\\
                              a,\ b \end{array}\right](z-1)^m
\sum_{k=0}^{\infty}\frac{(a+m)_k(b+m)_k}{k!(m+k)!}(1-z)^k\times
\nonumber\\
&{}&\qquad\times\bigl[\log(1-z)-\Psi(k+1)-\Psi(k+m+1)+\nonumber\\
&{}&\qquad\qquad
+\Psi(a+k+m)+\Psi(b+k+m)\bigr].
\label{e7}
\eea
Here $\Psi(x)$ is the logarithmic derivative of the $\Gamma$-function.
In the massless case, $m=1$ in (\ref{e7}).

Coefficients $C_{k,m}$ determine the boundary values of the field $\Phi$.
Then, for action (\ref{action}), we obtain
\bea
&{}&\int_0^{a}d\tau\int_0^{2\pi}d\vph\int_{\sin\th=0}^{\sin\th=1-\eps}d\th\,
\frac{\sin\th}{\cos\th}\left\{\d_\tau\Phi\d_\tau\Phi+\d_\th\Phi\d_\th\Phi
+\frac{1}{\sin^2\th}\d_\vph\Phi\d_\vph\Phi\right\}=\nonumber\\
&{}&\qquad=\int_0^a\,d\tau\int_0^{2\pi}\,d\vph \Phi(\tau,1-\eps,\vph)
\sin\th\left.\frac{\d}{\d\sin\th}
\Phi(\tau,\sin\th,\vph)\right|_{\sin\th=1-\eps}.
\label{e8}
\eea
Keeping only the logarithmically divergent and finite parts as $\eps\to0$
and using the standard formulas for $\Psi$-functions,
we obtain action (\ref{action})
in the form of the mode expansion ($C^*_{k,m}\equiv C_{-k,-m}$)~\cite{Ch1}
\bea
&{}&-\sum_{k,m\in{\bf Z}}^{}|C_{k,m}|^2\,
\left(\frac{m^2}{4}+w^2\right)
\bigl[\log\eps+\Psi\left(1+{m}/{2}+iw\right)
+\Psi\left(1+{m}/{2}-iw\right)+\bigr.\nonumber\\
&{}&\qquad\qquad\qquad
+\bigl. \Psi\left(1-{m}/{2}+iw\right)+
\Psi\left(1-{m}/{2}-iw\right)-4\Psi(1)\bigr],
\label{e9}
\eea
where $w$ and~$q$ are in (\ref{ww}), (\ref{e4}).

\newsection{Massive modes}

Including into consideration the ``internal'' (compact) degrees of
freedom (e.g., assuming the compact manifold to be a sphere $S^{d+1}$), 
we obtain an additional term in initial action (\ref{action}),
\bea
&{}&\int\frac{dx\,dy\,ds}{s^3}\bigl\{s^2\d_s\Phi\d_s\Phi+s^2(\d_x\Phi\d_x\Phi
+\d_y\Phi\d_y\Phi)+l(l+d)\Phi^2\bigr\}=\nonumber\\
&{}&\qquad=\int\,\tan\th d\tau\,d\th\,d\vph\left\{\d_\tau\Phi\d_\tau\Phi
+\d_\th\Phi\d_\th\Phi
+\frac{1}{\sin^2\th}\d_\vph\Phi\d_\vph\Phi
+\frac{l(l+d)}{\cos^2\th}\Phi^2\right\}.\label{action1}
\eea
Separating the variables as above, we obtain that the
restrictions on the ``torus'' coordinates~$\vph$ and~$\tau$  are as
in (\ref{vph}) 
and only the equation for the
$\th$-component is changed,
\be
\frac{\cos\th}{\sin\th}\d_\th\left(\frac{\sin\th}{\cos\th}
\d_\th Y_{k,m}\right)-\frac{m^2Y_{k,m}}{\sin^2\th}-\frac{w^2}{4}Y_{k,m}-
\frac{l(l+d)Y_{k,m}}{\cos^2\th}=0.
\label{sph1}
\ee
Introducing a new quantity $\rho$,
\be
4\rho(\rho-1)=l(l+d), \qquad \rho>0.
\ee
the solution to (\ref{sph1}) that is regular at $\th=0$ becomes
\be
Y_{k,m}(\th)=(\cos\th)^{2-2\rho}(\sin\th)^{|m|}
{}_2F_1\left(\frac{|m|}{2}+1-\rho+iw, \frac{|m|}{2}+1-\rho-iw; |m|+1;
\sin^2\th\right),
\label{Ykm1}
\ee
where again $w=-\frac{mb}{2a}+\frac{\pi k}{a}$.

Important particular case is $d=2$ (AdS$_3\times S^3$)
where $\rho=l/2+1$ and we must use
formula (\ref{e7}) in order to find the asymptotic
behavior. In this case, the nonlocality is again encoded
in the $\Psi$-function terms.

\newsection{Correlation function for massless modes}

Turn now to expression (\ref{e9}). The mode multiplier is the Fourier
transform of the correlation function of the boundary CFT field on 
the torus.
We are interested in the
nonlocal contribution to Green's functions on torus coming from this
formula and disregard all local terms, which can be removed by
a proper renormalizing procedure. (Therefore, we consider two-point
correlation functions rather than Green's functions for a two-dimensional
problem; the latter are distributions and necessarily demand local
contributions to be taken into account~\cite{AV}.)

The $m$-dependence in (\ref{e9}) is analytic~\cite{Ch1} and
we may represent the $\Psi$-function using the formula
$$
\Psi(1+a)-\Psi(1+b)=\sum_{j=0}^{\infty}\left(\frac{1}{1+j+b}
-\frac{1}{1+j+a}\right).
$$
Let us find the CFT correlation functions                        
that follow from (\ref{e9}).
The term
$\left(\frac{m^2}{4}+w^2\right)$ in front of the summand is nothing but
the Laplacian action on the Riemann surface, i.e.,
we obtain that the correlation function is
\bea
G(\tau,\vph)&=&
\frac{1}{4}\bigl(\d_\vph^2+\d_\tau^2\bigr)
\sum_{m,k\in \bf Z}^{}
\e^{im\vph+i\left[-\frac{mb}{a}+\frac{2\pi k}{a}\right]\tau}
\left[2\log\eps+\sum_{(\pm),\pm}
\Psi\bigl(1(\pm){m}/{2}\pm iw\bigr)-4\Psi(1)\right]=\nonumber\\
&=&
\frac{1}{4}\bigl(\d_\vph^2+\d_\tau^2\bigr)
\sum_{m,k\in \bf Z}^{}
\e^{im\vph+i\left[-\frac{mb}{a}+\frac{2\pi k}{a}\right]\tau}\times\nonumber\\
&{}&\qquad\times\left[2\log\eps+\sum_{(\pm),\pm}
\sum_{l=1}^{\infty}\left(\frac{4}{l}
-\frac{1}{l(\pm) \frac{m}{2}\pm i\bigl(-\frac{mb}{2a}
+\frac{\pi k}{a}\bigr)}\right)\right].
\label{d2}
\eea
The sign $\sum_{(\pm),\pm}$ in (\ref{d2}) denotes the sum over
four terms with all possible appearances of ``$+$'' and ``$-$'' signs.
We distinguish between two appearances of $\pm$ signs by
taking one of them in parentheses.
Constant and polynomial terms are irrelevant to our discussion as they
produce only local contributions.

First, we take the sum over $k$,
\be
\sum_{k=-\infty}^{\infty}{\e^{2\pi ik\tau/a}\over l(\pm) \frac{m}{2}\mp
\bigl(\frac{imb}{2a}-\frac{i\pi k}{a}\bigr)}=f(\tau/a).
\label{d3}
\ee
The function $f(\tau/a)$ is periodic under the shift
$\tau\to\tau+1$ and
satisfy the functional equation
\be
\pm\frac{1}{2a}\d_{\tau/a}f(\tau/a)+\left(l(\pm)\frac{m}{2}
\mp\frac{imb}{2a}\right)f(\tau/a)=\delta^1_{\Pi}(\tau/a),
\label{d4}
\ee
where $\delta^1_{\Pi}(x)$ is the periodic $\delta$-function with the
unit period. 
A (unique) solution to (\ref{d4}) that is periodic in $\tau$ is a
saw-tooth-like exponential curve
$$
f(\tau/a)=A\e^{\chi\tau/a}\quad\hbox{for}\quad \tau/a\in (0,1),
$$
which is to be continued periodically to the whole $\bf R$. Equation (\ref{d4})
gives
$$
\chi=\mp 2a\bigl(l(\pm){m}/{2}\bigr)+imb \quad\hbox{and}\quad
A=\frac{\pm 2a}{1-\e^{\mp2a\bigl(l(\pm)m/2\bigr)+imb}}.
$$
Therefore, the remaining nonlocal terms are combined into the sum
\be
-\sum_{{m\in {\bf Z}\atop l=1}}^{\infty}\e^{im\vph-imb\tau/a}
\sum_{\pm,(\pm)}^{}
\frac{\pm2a\e^{\mp 2\bigl(l(\pm){m}/{2}\bigr)\tau+imb\tau/a}}
{1-\e^{\mp2a\bigl(l(\pm)m/2\bigr)+imb}},\quad 0\le\tau<a.
\label{d5}
\ee
Expression (\ref{d5}) becomes
\be
\sum_{{m\in {\bf Z}\atop l=1}}^{\infty}\Re\left[
-2a\frac{\e^{mz+2il\vph}}{1-q^m\e^{2ilb}}
+2a\frac{\e^{mz-2il\vph}}{1-q^m\e^{-2ilb}}\right],
\qquad z\equiv\tau+i\vph.
\label{i1}
\ee
It is convenient to represent the denominators in (\ref{i1}) as
\be
\frac{1}{1-q^m\e^{i\chi}}=\left\{
\begin{array}{ll}
       \phantom{-}\sum_{p=0}^{\infty}q^{m\cdot p}\e^{ip\chi}&\quad
       \hbox{for}\quad |q|>1,\ m<0,\cr
       -\sum_{p=1}^{\infty}q^{-m\cdot p}\e^{-ip\chi}&\quad
       \hbox{for}\quad |q|>1,\ m>0.
\end{array}
\right.
\label{i2}
\ee
Now, it is easy to sum over~$l$ and~$m$ in (\ref{i1}) (the term with
$m=0$ vanishes). We obtain
\be
G(z,\ov z)=\frac{1}{4}\d_z\d_{\ov z}\Im\left\{
2a\sum_{p=1}^{\infty}\frac{1}{q^p\e^{-z}-1}\cot(\vph-bp)
-2a\sum_{p=0}^{\infty}\frac{1}{q^p\e^{z}-1}\cot(\vph-bp)
\right\}
\label{i3}
\ee
and it is straightforward to take the derivatives. The answer is
\be
G(z,\ov z)=\frac{a}{4}\sum_{p=-\infty}^{\infty}
\frac{1}{\sinh^2\frac{z+ap+ibp}{2}\sinh^2\frac{\ov z+ap-ibp}{2}}.
\label{Green}
\ee
This is the correlation function of two Yang--Mills tensor
field insertions on torus obtained in~\cite{Ch1} for $b=0$. 
At singularity points, it has the proper
behavior $G(r)\sim1/r^4$, which is character for the
weight two CFT fields. 

\newsection{Exact correlation function for the conical singularity case}

Using our technique, we are able to find 
the correlation functions also in the
case where a conical singularity appears at the axis $\theta=0$ of metric
(\ref{a2}). This corresponds to the following double periodic conditions
for the variable $z=\tau+i\vph$:
\be
z\to z+\frac{2\pi i}{a},\quad z\to z+\frac{2\pi}{ad}\xi,
\quad\hbox{where}\quad \xi=a+ib,\quad a,b,d\in {\bf R}.
\label{i5}
\ee
Actually, we can calculate the correlation function in the most general case
with torsion where $\tau$ variable is also shifted by the first 
transformation. Then, the two-dimensional eigenfunctions are
\be
\Phi_{k,l}(\tau,\vph)=\e^{i(al+ck)\vph+i(-lb+kd)\tau}
\label{i6}
\ee
(in (\ref{i5}), we set $c=0$) and the formula for the function 
$Y_{k,m}(\sin\theta)$ is exactly (\ref{e5}) where we must set
\be
m=al+kc,\qquad w=-lb+kd.
\label{i7}
\ee
Then, acting similarly to (\ref{d2})--(\ref{i2}), we obtain the expression
with the single remaining summation over~$n$,
\bea
&{}&G(z,\ov z)\sim \d_z\d_{\ov z} 2\Re\sum_{n=-\infty}^{\infty}
\biggl(\frac{i}{\pi\rho}
\frac{1}{\e^{z-\ov z\frac{\ov \rho}{\rho}+4\pi i\frac{n}{\rho}}-1}\cdot
\frac{1}{\e^{i\Delta\frac{\ov z}{\rho}+2\pi in\frac{\ov \xi}{\rho}}-1}
\biggr.
\nonumber\\
&{}&\qquad\qquad\qquad
+\biggl.\frac{i}{\pi\ov\rho}
\frac{1}{\e^{\ov z- z\frac{\rho}{\ov\rho}-4\pi i\frac{n}{\ov\rho}}-1}\cdot
\frac{1}{\e^{i\Delta\frac{z}{\ov\rho}+2\pi in\frac{\xi}{\ov\rho}}-1}
\biggr)\,,
\label{i8}
\eea
where $\xi\equiv a+ib$, \ $\rho\equiv c+id$, and $\Delta\equiv ad+bc$.

From now on, 
we restrict the consideration only to
the case without torsion, $c=0$. Also, we assume 
$a\in{\bf Z}_+$ to be a positive integer. Then, (\ref{i8}) becomes
$$
G(z,\ov z)=\frac{2\pi}{d}\d_z\d_{\ov z}\sum_{n=-\infty}^{\infty}
\left[\frac{1}{\e^{z+\ov z+\frac{4\pi n}{d}}-1}
\cdot\frac{\e^{a(z+\ov z)+\frac{4\pi na}{d}}-1}
{\Bigl(\e^{az+2\pi n\frac{\xi}{d}}-1\Bigr)
\Bigl(\e^{a\ov z+2\pi n\frac{\ov\xi}{d}}-1\Bigr)}\right].
$$
The term in brackets can be presented (for $a$ positive integer) as
$$
\frac{1}{a}\sum_{j=0}^{a-1}\frac{1}
{\Bigl(\e^{z+2\pi n\frac{\xi}{ad}+2\pi i\frac{j}{a}}-1\Bigr)
\Bigl(\e^{\ov z+2\pi n\frac{\ov\xi}{ad}-2\pi i\frac{j}{a}}-1\Bigr)},
$$
which eventually gives the answer for problem (\ref{action}) with 
periodic conditions (\ref{i5}) for $a\in{\bf Z}_+$:
\be
G(z,\ov z)=\frac{\pi}{2da}\sum_{n=-\infty}^{\infty}\sum_{j=0}^{a-1}
\frac{1}{\sinh^2\left(\frac{z}{2}+\pi n\frac{\xi}{ad}+\pi i\frac{j}{a}\right)
\sinh^2\left(\frac{\ov z}{2}+\pi n\frac{\ov\xi}{ad}-\pi i\frac{j}{a}\right)}
\label{i9}
\ee

The formula (\ref{i9}) resembles (\ref{Green}) with one, but important,
difference. We can write (\ref{Green}) as
\be
\sum_{p,m_1,m_2\in {\bf Z}}^{}\frac{4a}
{\bigl(z+(a+ib)p+2\pi im_1\bigr)^2\bigl(\ov z+(a-ib)p+2\pi im_2\bigr)^2}.
\label{i10}
\ee
Then, turning to (\ref{i9}), we obtain after rescaling $z\to z/a$,
\be
\sum_{p,m_1,m_2\in {\bf Z}}^{}\sum_{j=0}^{a-1}
\frac{8\pi a^3}{d}\frac{1}
{\bigl(z+2\pi\xi p+2\pi i(j+m_1a)\bigr)^2
\bigl(\ov z+2\pi\ov \xi p-2\pi i(j+m_2a)\bigr)^2}.  
\label{i11} 
\ee
In expression (\ref{i10}), the singularities in~$z$ and~$\ov z$ come always
at the same~$p$, but at all possible (independent) shifts in the angular
variable~$\vph$. Exrpession (\ref{i11}) enjoys the first of these properties
but angular shifts are no more independent---the mismatch is always an
multiple of~$a$ and, for instance, in the limit $a\to\infty$, the zero
mismatch term prevails, which automatically produces a $T$-dual answer.

\newsection{Discussion}

Having an AdS asymptotic metric in the boundary domain, we can continue it
in a bulk in (at least two) different ways: as an empty space or as a space
with the BTZ black hole singularity. From the CFT standpoint, these two
continuations are connected by the $T$-duality transformation; therefore, 
adopting a viewpoint that we must take a sum over all possible continuations
of the AdS metric (related, probably, to instantonic modes) for obtaining
a proper correlation function on the boundary, we obtain, instead of 
(\ref{Green}), the {\it sum} of (\ref{Green}) and its $T$-dual.

Obtained expression (\ref{Green}) describes the
two-point correlator $\langle G_{\mu\nu}^2(0,0)G_{\mu\nu}^2(z,\ov z)\rangle$
of the Yang--Mills tensor insertions
at nonzero temperature, which, from the conformal field theory standpoint,
corresponds to the case of a rectangular torus.
This demonstrates again that
the AdS/CFT correspondence holds in our case where no singularity at
the AdS time infinity is assumed. A more interesting (but far more
involved technically) is the problem of verifying this correspondence
in actual gravitational calculations of a multi-point correlation
functions.

The last problem to discuss is the $T$-duality under the modular
transformation $a\to1/a$ in different prescriptions. 
In the prescription of~\cite{Bonelli}.
it is possible to obtain the solid torus by factorizing the
AdS$_3$ space with respect to the action of the Abelian group generated
by two shifts (parabolic elements of $SL(2,\bf C)$): $z\to z+1$ and
$z\to z+i\xi$ (the third coordinate, $s$, is invariant).
Then, the fundamental domain is the half-infinite
cylinder $0<\Re z\le 1$, \ $0<\Im z\le \xi$, \ $\eps<s<M$
and, geometrically, this corresponds (in the three-dimensional picture) to
the closed torus inside which there is a singular (as $M\to\infty$)
torus of AdS radius
$\sim 1/M$ and length $\xi/M$. Then, in the limit
$\eps\to0$, \ $M\to\infty$, the corresponding temperature
correlation function becomes
$$
G_{Bon}(z,\ov z)\sim\sum_{n,m=-\infty}^{\infty}
\frac{1}{(z+n+i\xi m)^{2}(\ov z+n-i\xi m)^{2}},
$$
which apparently differs from (\ref{Green}) but coincides with
(\ref{i9}) in the limit $a\to\infty$ (with the proper rescaling
of $\vph$ and the Green's function). This, however, is not 
very surprising as
the case $a\to\infty$ corresponds to the case where the factorization
group is generated by two parabolic (oricyclic) elements of the 
$SL(2,{\bf C})$ group.

Already in the free field theory, there remain
questions on the mass spectrum, on the generalization 
to higher dimensions, etc. Of special interest is the question
how to consider solid Riemann surfaces of higher genera.
The construction works well in this case, but the
$\eps$-regularized surface, or the boundary of the integation domain,
cannot be described in the invariant
distance terms (the structure of the set
of closed geodesics becomes very involved already starting from genus two);
however, we hope that one can obtain a proper answer using
an approximation technique. In the present calculations, we
disregard all local contributions. However, these contributions
becomes important when considering additional boundary terms in the
initial action. It would be interesting to check whether the Hamiltonian
prescription of~\cite{ArFr} holds in this case.

It is also interesting to relate the obtained correlation functions
(\ref{Green}) and (\ref{i9}) with the $S$-matrix scattering problem
in AdS$_3$~\cite{Gid}.

\newsection{Acknowledgements}
The author thanks M.~A.~Olshanetsky for the valuable remark. The work
was supported by the Russian Foundation for Basic Research (Grant
No.\,98--01--00327).


\begin{thebibliography}{99}
\bibitem{Maldacena} J.~M.~Maldacena, {\sl Adv. Theor. Math. Phys.}
{\bf 2} (1998) 231--252 (hep-th/9711200).
\bibitem{Polyakov} S.~S.~Gubser, I.~R.~Klebanov, and A.~M.~Polyakov,
{\sl Phys. Lett.} {\bf B428} (1998) 105--114 (hep-th/9802109).
\bibitem{Witten} E.~Witten, {\sl Adv. Theor. Math. Phys.} {\bf 2} (1998)
253--291 (hep-th/9802150).
\bibitem{inter} W.~Mueck and K.~S.~Vishwanatan, {\sl Phys. Rev.} {\bf D58}
(1998) 41901 (hep-th/9804035);\\
Hong Liu and A.~A.~Tseytlin, ``{\it On four-point functions in the CFT/AdS
correspondence},'' hep-th/9807097;\\
D.~Z.~Freedman, S.~D.~Mathur, A.~Matusis, and L.~Rastelli, ``{\it Comments
on four-point functions in the CFT/AdS correspondence},'' hep-th/9808006;\\
E.~D'\,Hoker, 
D.~Z.~Freedman, S.~D.~Mathur, A.~Matusis, and L.~Rastelli, ``{\it Graviton
exchange and complete 4-point funcitons in the AdS/CFT 
correspondence},'' hep-th/9903196.
\bibitem{HF} E.~D'\,Hoker and D.~Z.~Freedman, ``{\it Gauge boson exchange
in AdS$_{d+1}$},'' hep-th/9809179.
\bibitem{Ch1} L.~Chekhov, ``{\it AdS/CFT correspondence on torus},''
hep-th/9811146.
\bibitem{Bonelli} G.~Bonelli, ``{\it Holography and CFT on generic
manifolds},'' hep-th/9810194.
\bibitem{Wit2} E.~Witten, ``{\it Anti-de~Sitter space, thermal phase
transition, and confinement in gauge theories},'' hep-th/9803131.
\bibitem{HP83} S.~W.~Hawking and D.~Page, {\sl Commun. Math. Phys.}
{\bf 87} (1983) 577--588.
\bibitem{CT} S.~Carlip and C.~Teitelboim, {\sl Phys. Rev.} {\bf D51}
(1995) 622--632.
\bibitem{MS} J.~Maldacena and A.~Strominger,
``{\it AdS$_3$ black holes and a stringy exclusion principle},''
hep-th/9804085.
\bibitem{Manin} D.~Hejhal, {\sl Adv. Math.} {\bf 15} (1975) 133--156.
\bibitem{Sel} W.~P.~Thurston, ``The geometry and topology of
{\rm3}-manifolds,'' Princeton notes, 1979.
\bibitem{BPM} A.~P.~Prudnikov,
Yu.~A.~Brychkov, and O.~I.~Marichev, Integrals and Series.
Vol.~3. Additional Chapters. Nauka, Moscow, 1985 [in Russian].
\bibitem{AV} I.~Ya.~Arefeva and I.~V.~Volovich,
{\sl Phys. Lett.} {\bf B433} (1998) 49--55.
\bibitem{ArFr} G.~E.~Arutyunov and S.~A.~Frolov, ``{\it On the origin of
supergravity boundary terms in the AdS/CFT correspondence},''
hep-th/9806216.
\bibitem{Gid} S.~B.~Giddings, ``{\it The boundary $S$-matrix and the AdS
to CFT dictionary},'' hep-th/9903048.
\end{thebibliography}
\end{document}